\documentstyle[preprint,aps]{revtex}

\begin{document}


\title{Charge and Statistics of Quasiholes 
in Pfaffian States of Composite Fermion 
Excitations}

\author{Piotr Sitko} 

\address{Institute of Physics, 
Wroc\l{}aw University of Technology,\\ Wybrze\.ze Wyspia\'{n}skiego
 27, 50-370 Wroc\l{}aw, Poland.}

\maketitle

\begin{abstract}
The charge of quasiparticles in Pfaffian 
states of composite fermion excitations (the presence of which is 
indicated by recent experiments) 
is found. 
At the filling fraction of the Pfaffian state
$\nu=p/q$ (of the lowest Landau level) the charge is $\pm e/(2q)$.
As in the case of the Pfaffian state
of electrons the statistics of $N_{qh}$ quasiholes in the Pfaffian state
corresponds to
the spinor representation of  $U(1)\times SO(2N_{qh})$
(the continuous extension of the braid group).
Here $U(1)$ is given by the phase factor
$e^{i(\frac{1}{8}+\frac{1}{4m})\pi}$ with $m=1+\alpha$, $\alpha$ -- the
exclusion statistics parameter of Jain quasiparticles.
The possiblity of Read-Rezayi states of Jain quasiparticles is also
discussed.
\end{abstract}


Recently, the non-Laughlin and non-Jain states were observed in the
fractional quantum Hall effect \cite{Pan}. 
They can be interpreted as condensed
states of composite fermion excitations \cite{cfhierarchy,LF}.
Even-denominator states (i. e. at filling fractions with an even
denominator) can be seen as Pfaffian states of Jain quasiparticles.
Odd-denominator states correspond to the condensed states of composite
fermion excitations which can either be Laughlin \cite{Laughlin}
(or Jain) states
or condensed states of different origin. One possiblity 
(as competing to Laughlin-Jain states) is
the class of states proposed by Read and Rezayi \cite{ReadRezayi} 
for
electron states in higher Landau levels
(Read-Rezayi states result from clustering of a number of particles,
the Pfaffian state is the state of pairs of particles \cite{Moore}).
The other proposed states for the origin of odd-denominator states
are Halperin paired states
\cite{Halperin,RR2,Quinn}.

The composite fermion approach predicts  the so-called Jain states 
at  filling fractions of
the form $\nu=\frac{n_{0}}{2p_{0}n_{0}+\beta_{0}}$ 
(where $2p_{0}$ -- the Chern-Simons composite fermion parameter being an even
number,
$n_{0}$ -- the number of effective shells filled,
$\beta_{0}$ -- the sign of the effective 
field with respect to the
external magnetic field).
The effective field is found when the mean Chern-Simons composite fermion
field is added to the external magnetic field.
Let us introduce the condensed states of composite fermion
excitations (following \cite{cfhierarchy}) which appear at the filling
fraction of the form:
\begin{equation}
\label{fraction}
\nu_{e}^{-1}=2p_{0}+\frac{\beta_{0}}{n_{0}+\nu_{1}}
\end{equation}
where $\nu_{1}$ is the fraction in which the  $(n_{0}+1)$-th 
effective shell is partially filled.
All recently observed states \cite{Pan} can be described within (1)
(e. g. $\nu_{e}=4/11$, with $p_{0}=1$, $n_{0}=1$, $\beta_{0}=1$,
$\nu_{1}=1/3$).
The states $4/13$, $5/17$, correspond  to the $\beta_{0}=-1$ case
(the effective field is opposite to the external magnetic field).
The state $7/11$ can either be seen as the $\beta_{0}=-1$ state
(with $n_{0}=2$) or as the $4/11$ state of holes conjugated to electrons.
The partial fillings of interest (as observed in the experiment) 
are $\nu_{1}=1/3$, $2/3$, and
$1/5$. Even-denominator states correspond to $\nu_{1}=1/2$.
In this paper we are also going to refer to Read-Rezayi states defined at
filling fractions (in the fermion representation of fractional statistics
particles):
\begin{equation}
\nu_{1}=\frac{k}{2+k}\; 
\end{equation}
($k=2$ for the Pfaffian, for $k=1$ the Laughlin $1/3$ state is found).
All observed hierarchical states would correspond to even-$k$ Read-Rezayi
states,
i. e. $\nu_{1}=1/2$ (the Pfaffian $k=2$),
$\nu_{1}=2/3$ (or by the particle-hole conjugation $\nu_{1}=1/3$) --
$k=4$, $\nu_{1}=4/5$ (by the particle-hole conjugation $\nu_{1}=1/5$) --
$k=8$ (which correponds to the $\nu=6/17$ state).
In the last case it seems that the Laughlin $1/5$ state should rather be
preferred over the Read-Rezayi clusters of $8$ quasiparticles.
Assuming the possibility of Read-Rezayi states the question arises why only these three values of $k$ would be
of interest.
If one would have  the $k=4$ Read-Rezayi state, 
why a $k=3$ state would not appear.
Even worse, why the $k=8$ state if there would be no $k=5,6,7$ states.
(The case of $k=6$ would correspond to the $\nu_{1}=1/4$ Pfaffian state 
via particle-hole conjugation, if observed).
The simple test for Read-Rezayi states would be for example
the observation
of the $\nu_{1}=3/7$ state of composite fermion excitations
(it is the first Jain state which can not be represented as a Read-Rezayi
state).
However, we should stress that 
the Pfaffian state of composite fermion excitations is seen as the best
candidate  
for recently discovered even-denominator hierarchical states
$3/8$ and $3/10$.

It is worth to notice that the observed $3/8$ state corresponds to the
half-filling of Laughlin quasielectron shell (Laughlin quasielectrons of
the $1/3$ state).
However, the $3/10$ state  corresponds to the $1/4$ 
filling of the Laughlin quasihole shell. 
It is perhaps better to see the $3/10$ state as
the half-filled
quasielectron shell, but seen within the opposite effective field picture
(or, equivalently, to the half-filled quasihole level of quasiholes of the
$2/7$ state).
Within the composite fermion picture then, both  states ($3/8$
and $3/10$) correspond to a half-filling of the first excited effective
Landau level. It is known that the Pfaffian state of electrons is proposed
for the $\nu=5/2$ quantum Hall state (electrons in the first excited 
Landau
level, two underlying levels are occupied
by electrons of opposite spins). Also, the Read-Rezayi states are 
proposed for electrons in the
first excited (or higher)  Landau level.
Hence, it is interesting to consider also 
Read-Rezayi states of quasiparticles
as natural candidates (competing with the picture of Laughlin or Jain
states of quasiparticles).
Maybe, except for the $\nu=6/17$ state (the $1/5$ state of quasielectrons, or
the $4/5$ state of conjugated holes -- the Read-Rezayi state for $k=8$ --
$8$ holes make a cluster -- as discussed above).

At present in the case of odd-denominator
states the first guess aims at
 Laughlin or Jain  states of quasiparticles
\cite{cfhierarchy,LF}. One possible competing mechanism of incompressibility
may be due to  clustering of
quasiparticles as discussed by Read and Rezayi (in the case of electrons)
\cite{ReadRezayi}.
Laughlin (or Jain) states of Jain quasiparticles
are not seen in numerical studies (e. g. in
spherical systems \cite{QHS}) but the studies are far from being conclusive (the
number of quasiparticles is very low when exact diagonalization of
electrons is considered, usually 3-4 quasiparticles \cite{cfhierarchy,QHS}). 
The approximate numerical methods for higher numbers of
quasiparticles are also proposed \cite{cfhierarchy,MandalJain,Jain,Quinn}
but their 
accuracy seems to be difficult to
estimate.

There are examples in numerical studies (on a sphere)
which support the Pfaffian states of composite fermion excitations
(which should correspond to the $3/8$ states), e. g.
the states with the $L=0$ ground state:
$N_{e}=14$ $2S=33$ ($2S_{qe}=9$, $N_{qe}=6$) \cite{MandalJain},
$N_{e}=12$ $2S=29$ ($2S_{qe}=9$, $N_{qe}=4$) \cite{cfhierarchy,Jain}
($N_{e}$ electrons are in the angular momentum shell $S$, $N_{qe}$
quasielectrons
are in the effective shell $S_{qe}$).
The former can be seen as the Pfaffian state of quasielectrons
($2S_{qe}=2N_{qe}-3$) \cite{GWW}, the latter is  related by the
particle-hole conjugation (in the quasielectron shell).
There are also approximate results for a higher number of quasiparticles
pointing at a Pfaffian state (e. g.  for $N_{qe}=10$, $2S_{qe}=17$
in \cite{Quinn}).

We are going to define the charge of quasiparticles in Pfaffian (and
Read-Rezayi) states of composite fermion excitations.
Note that as Jain 
quasiparticles are anyons they can be seen
in the fermion (or boson) representation (with the appropriate
Chern-Simons field included).
We use the strong-pairing (strong-clustering) limit
which is valid   if  charge of excitations is considered
\cite{GWW,ReadRezayi,RR2}.
The (non-Abelian) statistics of quasiholes in Pfaffian states
corresponds to the spinor representation of  $U(1)\times
SO(2N_{qh})$ ($N_{qh}$ -- a number of quasiholes) of the dimension
$2^{N_{qh}-1}$. In the present case one can consider
Pfaffian states  of Jain quasiparticles  in the   boson
representation.
It means we first consider the Pfaffian state of bosons,
and next we introduce the Chern-Simons field appropriate 
for the fractional 
statistics.
Such operation affects only the $U(1)$ term of the representation
(Fradkin {\it et al.} \cite{FNTW} discussed the similar case
where the Pfaffian fermion state is found within the boson representation
of fermions).
The Pfaffian state (in the plane, $z_{i}=x_{i}+iy_{i}$) is defined as:
\begin{equation}
\label{pfaffian}
\Psi_{Pf}=Pf(\frac{1}{z_{i}-z_{j}})\prod_{i>j}(z_{i}-z_{j})^{m}
e^{-\frac{\sum|z_{i}|^{2}}{4a_{0}^{2}}}\; ,
\end{equation}
$a_{0}$ is the magnetic length,
$m$ is odd for bosons, and is even for fermions
(at the respective filling fraction $1/m$).
One can notice that $m$ appears only in the Laughlin-like part of the
wave function (\ref{pfaffian}). For four quasiholes in the Pfaffian state
the braiding matrix is \cite{FNTW}
\begin{equation}
\label{braid}
\frac{e^{i\pi(\frac{1}{8}+\frac{1}{4m})}}{\sqrt{2}}
\left(\begin{array}{cc} 1&1\\ -1&1
\end{array}\right)
\end{equation}
(in the spinor representation).
We note that the second term in the phase factor ($\frac{\pi}{4m}$)
can be found in the strong-pairing limit \cite{GWW,Nayak,WenLee}.
Hence, the value of quasihole charge is $\frac{1}{4m}Q_{c}$ where $Q_{c}$ is
the charge of pairs of particles, for $Q_{c}=2e$ one gets the charge
$\frac{e}{2m}$.
The Pfaffian state of anyons (Jain quasiparticles)
corresponds to a noninteger $m$ \cite{GWW} 
(which describes the statistics),
and we define $m=1+\alpha$ where $\alpha$ is the exclusion  statistics parameter 
of anyons \cite{hierarchy} 
(in the magnetic field, at the effective filling $1/2$, in the
fermion representation, $\alpha\ge 0$).

We discuss below the strong-pairing (strong-clustering) limit.
Let us consider the filling fraction of the form 
\begin{equation}
\nu_{e}^{-1}=2p_{0}+\frac{\beta_{0}}{1+\nu_{1}}
\end{equation}
$\nu_{1}$ is the filling fraction of composite fermion excitations in the
effective $n_{0}=2$ shell.
We assume Read-Rezayi states of quasielectrons (for the Pfaffian $k=2$), 
hence \cite{ReadRezayi}
\begin{equation}
\nu_{1}^{-1}=1+\frac{2}{k}\; .
\end{equation}
The total flux entering the system is
\begin{equation}
\Phi^{ex}=N_{e}(\nu_{e}^{-1})\frac{hc}{e}\; .
\end{equation}
One gets $N_{e}=\frac{1+\nu_{1}}{\nu_{1}}N_{qe}$ ,
$Q_{qe}=e/q_{p}$
(quasielectrons partially fill the first excited effective shell),
$q_{p}$ is the denominator  of the parent state \cite{Haldane}.
Since $k$ quasielectrons make a cluster one gets
$N_{c}=\frac{N_{qe}}{k}$ (the number of clusters), $Q_{c}=kQ_{qe}$ ,
and 
\begin{equation}
\Phi^{ex}=\nu_{e}^{-1}\frac{1+\nu_{1}}{\nu_{1}}k^{2}\frac{1}{q_{p}}N_{c}
\frac{hc}{Q_{c}} \; .
\end{equation}
In the case of $N_{c}$ anyons in the Laughlin state 
\cite{hierarchy} (of charge $Q_{c}$ and the statistics 
$\theta/\pi=(\alpha_{str}^{qh})^{-1}(mod\; 2)$)
\begin{equation}
\Phi^{ex}=(\alpha_{str}^{qh})^{-1}\frac{hc}{Q_{c}}N_{c}
\end{equation}
with quasihole excitations of the exclusion statistics
$\alpha_{str}^{qh}$, and the quasihole charge $Q=\alpha_{str}^{qh}Q_{c}$.
We assume that $k$-clusters condense at a Laughlin state,
then the excitations charge $Q$ is (if absolute values of charges are
considered)
\begin{equation}
Q^{-1}
=\nu_{e}^{-1}\frac{1+\nu_{1}}{\nu_{1}}k^{2}\frac{1}{q_{p}}\frac{1}{Q_{c}}
\end{equation}
\begin{equation}
Q^{-1}e=\nu_{e}^{-1}\frac{1+\nu_{1}}{\nu_{1}}k\; .
\end{equation}
One gets then
\begin{equation}
eQ^{-1}=4p_{0}+2\beta_{0}+k(4p_{0}+\beta_{0})
\end{equation}
and
\begin{equation}
\nu_{e}=\frac{2+2k}{eQ^{-1}}\; .
\end{equation}
For $k$  even the filling fraction can be simplified
(the numerator and the denominator in $\nu_{e}$ may be divided by two).
Hence, the charge of excitations  equals $\pm e/q$
for $k$ odd
and $\pm e/(2q)$ for $k$ even.
The same conclusion can be reached at the filling fraction of the form:
\begin{equation}
\label{Jainqh}
\nu_{e}^{-1}=2p_{0}+\frac{\beta_{0}}{2-\nu_{1}}
\end{equation}
where holes conjugated to quasielectrons are considered
(Jain quasiholes of the  state $\nu_{e}^{-1}=2p_{0}+\frac{\beta_{0}}{2}$).
In that case
\begin{equation}
eQ^{-1}=k(2p_{0}+\beta_{0})+8p_{0}+2\beta_{0}
\end{equation}
and 
\begin{equation}
\nu_{e}=\frac{4+k}{eQ^{-1}}\; .
\end{equation}
The charge of quasiparticles in fractional quantum Hall systems
can be detected, e. g. in "shot-noise" experiments
\cite{Chamon,Picciotto,Seminadayar,Reznikov}. 

The statistics of excitations in the strong-pairing limit
can also  be determined.
Let us consider the Pfaffian, i. e. $k=2$, then
\begin{equation}
eQ^{-1}=4(3p_{0}+\beta_{0})= [\alpha^{qh}_{str} Q_{c}]^{-1}e\; ,
\end{equation}
$\alpha^{qh}_{str}$ is the exclusion statistics parameter of quasiholes
in the strong-pairing limit.
When one considers the statistics of quasiholes within the strong-pairing
limit one finds
\begin{equation}
\alpha^{qh}_{str}=\frac{1}{Q_{c}Q^{-1}}=\frac{q_{p}}{2eQ^{-1}}\; ,
\end{equation}
$q_{p}=2p_{0}n_{0}+\beta_{0}$ -- the denominator of the parent Jain state
(here $n_{0}=1$).
And
\begin{equation}
\alpha^{qh}_{str}=
\frac{\beta_{0}+2p_{0}}{2\cdot 4(3p_{0}+\beta_{0})}\; .
\end{equation}
Let us assume that $\alpha_{qe}=m-1$,
then $m= \alpha_{qe}+1=\frac{2(\beta_{0}+3p_{0})}{\beta_{0}+2p_{0}}$
\cite{hierarchy}.
Then
\begin{equation}
\label{4m}
\alpha^{qh}_{str}=\frac{1}{4m}\; .
\end{equation}
The same relation can be found for Jain quasiholes in (\ref{Jainqh}).
The result (\ref{4m}) is the one which appears in 
the expression for the statistics
of excitations  in the Pfaffian state (\ref{braid}).
The state $3/8$ corresponds to the Pfaffian state of $\alpha_{qe}=5/3$
quasielectrons \cite{hierarchy}.
Then the phase factor in (\ref{braid}) is
\begin{equation}
\frac{\pi}{8}+\frac{3}{32}\pi \;.
\end{equation}
When one considers the Pfaffian state of 
the conjugated quasiholes (of the $2/5$ state,
$\alpha_{qh}=3/5$) one gets
$\frac{\pi}{8}+\frac{5}{32}\pi$.
Note, however, that
both cases correspond to opposite change in the external magnetic
field (e. g. quasihole-like, quasielectron-like excitations).
At $\nu=3/10$ we have $\alpha_{qe}=7/3$ \cite{hierarchy},
so that the phase factor is $\pi/8 +(3/40)\pi$
(or if one starts from $\nu=2/7$, $\alpha_{qh}=3/7$, then
one gets the phase $\pi/8+(7/40)\pi$).

In conclusion we found the charge of quasiparticles in condensed states of
composite fermion excitations seen as the Pfaffian and Read-Rezayi states. 
In the standard hierarchy
(e. g. the Haldane hierarchy \cite{Haldane} 
or the hierarchy of composite fermion excitations\cite{cfhierarchy})
the charge is always $\pm e/q$ where $q$ is the denominator of the filling
fraction ($\nu=p/q$; $p$, $q$ are coprime).
However, for Pfaffian or Read-Rezayi  states (where $k$ particles
make a cluster, $k=2$ for the Pfaffian) one gets $\pm e/q$ for $k$ odd
and $\pm e/(2q)$ for $k$ even (e. g. for the Pfaffian state).
Then, the charge of excitations in the proposed 
Pfaffian states $3/8$ and $3/10$
should be $\pm e/16$ and $\pm e/20$, respectively.
This provides also the possible identification  of Read-Rezayi
states of Jain quasiparticles (for even $k$)  
in charge-detecting experiments (e. g. in
"shot-noise" experiments).
For example the charge of excitations at $\nu_{e}=5/13$  (and
$\nu=4/11$) should be $\pm e/26$ ($\pm e/22$ 
for $\nu_{e}=4/11$) if one deals with
Read-Rezayi states of quasiparticles.
Hence, for all observed  states it should be possible to identify
Pfaffian and $k$-even Read-Rezayi states by future 
charge-detecting experiments.
The Pfaffian state can be further tested  by detecting (if possible)
the non-abelian
statistics of excitations.
The non-Abelian statistics of quasiholes  in Pfaffian states of Jain
quasiparticles is given
as the spinor representation of $U(1)\times SO(2N_{qh})$
(the continuous extension of the braid group) with $U(1)$
given by $e^{i(\frac{\pi}{8}+\frac{\pi}{4m})}$ where $m=1+\alpha$,
$\alpha$ -- the exclusion statistics of Jain quasiparticles
(at the effective filling fraction $1/2$ in the fermion representation).
Let us note that also excitations in Read-Rezayi states of Jain
quasiparticles (if observed) would obey non-abelian statistics.

This work was supported by the Polish Ministry of Scientific Research and
Information Technology,  grants No. 1 P03B 09826 and No. PBZ-MIN-008/P03/2003 




\end{document}